\documentclass[conference]{IEEEtran}
\usepackage{cite}
\usepackage{url}
\usepackage[hidelinks]{hyperref}
\usepackage{amsmath}
\usepackage[utf8]{inputenc}
\usepackage[utf8]{inputenc}
\usepackage{newunicodechar}
\newunicodechar{）}{)}
\newunicodechar{（}{(}

\usepackage{makecell}

\usepackage{tabularx}
\usepackage{array} 
\usepackage{booktabs}
\newcolumntype{L}{>{\raggedright\arraybackslash}X} 

\begin{document}

\title{Element and Everything Tokens: Two-Tier Architecture for Mobilizing Alternative Assets}

\author{\IEEEauthorblockN{Ailiya Borjigin\IEEEauthorrefmark{1}, Cong He\IEEEauthorrefmark{1}, Charles CC Lee\IEEEauthorrefmark{2}, Wei Zhou\IEEEauthorrefmark{1}}
\IEEEauthorblockA{\IEEEauthorrefmark{1}ProAI Laboratory, Probe Group, Singapore\\ Email: \{Ailiya, Cong\_He, Zhou\}@probe-group.com}
\IEEEauthorblockA{\IEEEauthorrefmark{2}Centre for Sustainable Development, University of Newcastle （Australia）, Singapore\\ Email: charles.cc.lee@newcastle.edu.au}}

\maketitle

\begin{abstract}
Alternative assets such as mines, power plants, or infrastructure projects are often large, heterogeneous bundles of resources, rights, and outputs whose value is difficult to trade or fractionalize under traditional frameworks. This paper proposes a novel two-tier tokenization architecture to enhance the liquidity and transparency of such complex assets. We introduce the concepts of \emph{Element Tokens} and \emph{Everything Tokens}: elemental tokens represent standardized, fully collateralized components of an asset (e.g., outputs, rights, or credits), while an everything token represents the entire asset as a fixed combination of those elements. The architecture enables both fine-grained partial ownership and integrated whole-asset ownership through a system of two-way convertibility. We detail the design and mechanics of this system, including an arbitrage mechanism that keeps the price of the composite token aligned with the net asset value of its constituents. Through illustrative examples in the energy and industrial sectors, we demonstrate that our approach allows previously illiquid, high-value projects to be fractionalized and traded akin to stocks or exchange-traded funds (ETFs). We discuss the benefits for investors and asset owners, such as lower entry barriers, improved price discovery, and flexible financing, as well as the considerations for implementation and regulation. 
\end{abstract}


\section{Introduction}
Programmable asset issuance---the process of creating digital representations of alternative assets on a blockchain---has been gaining momentum as a means to democratize investment and unlock liquidity in traditionally illiquid asset classes\cite{Waliczek2025,Joshi2022}. By enabling \emph{fractional ownership}, tokenization allows high-value assets like real estate, infrastructure, or commodities to be divided into smaller, more affordable units that can be bought and sold by a broader range of investors\cite{Waliczek2025}. For example, instead of requiring a single buyer for an entire commercial building or solar farm, tokenization makes it possible to distribute ownership among many investors, each holding a digital token that confers a portion of the asset's value or returns\cite{Waliczek2025}. This has the potential to lower entry barriers and increase market liquidity for assets that were once the exclusive domain of large institutions or ultra-high-net-worth individuals.

However, most existing tokenization approaches treat the asset as a monolithic whole represented by a single token, which can obscure the diverse components that contribute to its value and risk profile. Large-scale real assets are inherently \emph{heterogeneous}, often comprising multiple distinct elements—such as physical outputs, land or resource rights, operational permits, revenue streams, and environmental credits. In traditional markets, these components are bundled together, making the asset difficult to value transparently and to trade in parts. As a result, investors face high buy-in thresholds and cannot easily tailor their exposure to specific aspects of the asset\cite{Joshi2022}. It is, for instance, typically impossible to invest in just the future gold output of a mining project or only the energy production of a power plant without purchasing the entire asset or entering complex contracts\cite{Joshi2022}. This bundling leads to illiquidity and under-utilization of the asset’s potential value.

Existing financial innovations provide partial analogies but still fall short for heterogeneous real assets. Securitization and real estate investment trusts (REITs) fractionalize assets into shares, but these shares still represent a claim on the entire pooled asset or cash flow, not on specific sub-components\cite{FSB2024}. Similarly, basket tokens in decentralized finance (DeFi) and exchange-traded funds (ETFs) allow investors to hold a portfolio of assets through one token, but these typically bundle separate assets of a similar type (e.g., a collection of equities or cryptocurrencies) rather than decomposing a single complex asset into its elements. Moreover, traditional ETFs rely on authorized participants to perform creation and redemption in-kind to keep prices aligned with underlying asset value\cite{Todorov2021}. Security token offerings (STOs) for private assets have to date largely adopted a one-token-one-asset model, which does not provide internal divisibility by asset component. Even emerging concepts like fractional non-fungible tokens (NFTs) focus on dividing ownership shares of a singular item and do not natively handle assets composed of multiple revenue streams or rights.

Notably, recent research has begun to explore multi-token representations for single assets. For example, Joshi and Choudhury tokenized real estate by issuing multiple token types on an Ethereum ERC-1155 contract: unique tokens for distinct property rights and fungible tokens representing fractional shares associated with those rights\cite{Joshi2022}. This demonstrates the feasibility of a more granular approach to tokenizing asset components. Building on this insight, we propose a generalized two-tier token model that standardizes the decomposition of any complex asset into elemental parts, while also providing a mechanism to recombine those parts into a whole asset token for holistic trading and investment.

In this paper, we introduce the \textbf{Element and Everything Token} architecture and argue that it can significantly improve liquidity and price discovery for complex Alternative Assets. An \emph{Element Token} refers to a standardized, fungible token representing a specific, independently measurable component of an asset (for example, one ton of mineral output, one square meter of land usage rights for one year, one megawatt-hour of energy production, or one unit of carbon credit). These element tokens are designed to be fully collateralized by the underlying asset components and potentially fungible across projects (e.g., carbon credit tokens from different projects might be fungible if they adhere to the same standard). An \emph{Everything Token}, by contrast, represents an entire asset or project; it is constructed as a fixed bundle of various element tokens in proportions that reflect the asset's composition. The everything token can be thought of as analogous to a basket or ETF share that holds a predetermined mix of underlying tokens corresponding to the asset's constituent parts.

The remainder of this paper is organized as follows. Section II formalizes the element/everything token architecture, detailing the token design, conversion mechanics, and pricing relationships. Section III provides illustrative examples of how diverse assets—from mining and energy projects to industrial facilities and environmental assets—can be tokenized using this framework. Section IV discusses the practical implications, benefits, and challenges of implementing the architecture, including perspectives for investors, asset owners, and regulators. Section V concludes with a summary of findings and suggestions for future work in this domain.

\section{Literature Review}
\subsection{Tokenization of Real-World Assets}
Empirical studies of security token offerings (STOs) in real estate document that tokenization expands investor access and fractional ownership while secondary trading remains thin and influenced by crypto-specific frictions (fees, sentiment)\cite{Kreppmeier2023}. Broader surveys and policy analyses argue that tokenization can reduce issuance and transfer frictions, improve settlement finality, and mobilize new investor bases for infrastructure and sustainable projects\cite{Uzsoki2019,Tian2022,Tanveer2025}. These findings motivate architectures that (i) standardize claims at a granular level and (ii) design market microstructure to support continuous price discovery across heterogeneous claims.

\subsection{Exchange-Traded Funds and Creation/Redemption}
The ETF literature provides the closest analogue to our bidirectional convertibility. Classic and contemporary studies show that authorized participants (APs) arbitrage deviations between ETF prices and the value of their underlying baskets via primary-market creations and redemptions\cite{Madhavan2016}. When underlying assets are illiquid (e.g., corporate bonds), AP balance-sheet frictions can weaken the arbitrage link and allow persistent premiums/discounts\cite{PanZeng2017}. Flow-induced, non-fundamental demand in the ETF primary market can also predict returns and generate cross-sectional mispricing\cite{Brown2021}. These results inform our mechanism design: creation/redemption should be (a) programmatic and permissionless where compliant, (b) fee-calibrated to avoid abuse, and (c) supported by liquidity in both composite and element markets to keep prices near basket NAV.

\subsection{SPACs and Asset-Level Financing}
SPACs finance acquisitions using a shell equity that later merges with a target. Recent evidence highlights agency problems and costs relative to IPOs but also clarifies conditions under which SPACs offer flexibility\cite{GahngRitterZhang2023,HuangRitter2023}. Our \emph{Everything Token} plays an analogous role as a project-level financing shell, but with transparent, on-chain claims to standardized components and programmable cash-flow routing—bridging insights from SPAC financing with on-chain composability.

\subsection{DeFi Market Design: AMMs and Basket Construction}
DeFi advances formalize constant-function market makers (CFMMs), deriving no-arbitrage bands, price bounds, and multi-asset trade optimality via convex analysis\cite{AngerisAFT2020,AngerisCFMM2021,AngerisGeometry2023}. Analyses of Uniswap-type markets suggest that, under mild conditions, CFMM prices track external references, enabling robust on-chain price discovery for fungible tokens\cite{AngerisUniswap2019}. Multi-asset CFMMs provide natural liquidity backbones for element baskets. We leverage these results to design dual venues: (1) spot CFMMs for each element token and (2) a creation/redemption contract for the composite, so that arbitrage links micro and macro prices.

\subsection{Data Integrity and Oracle Mechanisms}
Bridging on-chain markets to off-chain realities requires secure oracle designs. Authenticated feeds such as Town Crier use trusted execution to attest to HTTPS data for smart contracts\cite{TownCrier2016}. Cryptographic protocols like DECO enable zero-knowledge proofs of statements about TLS sessions without trusted hardware\cite{DECO2019}. Recent software-engineering work emphasizes verifying oracle deviation impacts on DeFi protocols\cite{OVer2024}. For RWA tokenization, these lines motivate layered verification—legal attestations, IoT telemetry, and oracle proofs—to ensure that element tokens remain fully collateralized and auditable.

\section{Element/Everything Token Architecture}
\subsection{Design Overview}
In the proposed architecture, each complex asset is represented by a set of $n$ element tokens $E_1, E_2, \ldots, E_n$ and a single everything token $W$. Each element token $E_i$ corresponds to the smallest unit of a particular asset component that can be independently priced and traded. These could be physical quantities (commodities or outputs), rights or entitlements (land or spectrum usage rights, licenses, quotas), or financial derivatives of asset performance (e.g., a token for a portion of revenue or a carbon offset credit). The element tokens are \emph{standardized} and backed one-to-one by the specified asset components, ensuring that holding an element token entitles the bearer to a well-defined claim on the underlying asset or its cash flows. By isolating components, element tokens facilitate transparent price discovery for each aspect of the asset; for instance, the market can establish a price per MWh of solar electricity or per ton of copper from a given project.

The everything token $W$ is then defined as a fixed-proportion bundle of the $n$ element tokens that together constitute one whole asset unit. If one physical unit of the asset (e.g., one entire power plant or one mining operation) intrinsically includes $a_1$ units of component 1, $a_2$ units of component 2, ..., $a_n$ units of component $n$, then one everything token is composed of $(a_1 \, E_1,\, a_2 \, E_2,\, \ldots,\, a_n \, E_n)$. In other words, $W$ is a composite token representing the complete asset, such that:
\begin{equation}
    W \equiv a_{1} E_{1} + a_{2} E_{2} + \cdots + a_{n} E_{n},
\end{equation}
where the addition denotes aggregation of the bundle of element tokens. In practice, a smart contract could enforce this relationship by only allowing $W$ to be minted (created) when the required number of each $E_i$ is deposited (or locked) into a reserve, and conversely, allowing anyone to redeem one $W$ token to retrieve the underlying basket of $a_1$ of $E_1$, $a_2$ of $E_2$, ..., $a_n$ of $E_n$.

The fixed ratios $a_i$ are determined based on the asset's technical or legal structure. For example, if a solar farm project is defined by 100 MWh/year of energy production capacity, 1000 square meters of land lease, and 100 carbon credits per year, then an everything token for that project might be composed of those quantities of the respective element tokens. These proportions remain fixed for that project token; if the asset's capacity or entitlements change, the token specification would be updated or a new token series issued to reflect the new composition.

A critical feature of the architecture is that it supports \textbf{two-way convertibility} between the everything token and the element tokens. This means investors have the flexibility to either hold or trade the entire asset via $W$, or to disaggregate their position into individual $E_i$ tokens and possibly trade those separately. The convertibility is enabled by smart contracts that implement a \emph{bundle swap}: one $W$ can be exchanged for $(a_1 E_1, \ldots, a_n E_n)$, and conversely, anyone holding the exact bundle of $(a_1, \ldots, a_n)$ units of $E_1,\ldots,E_n$ can combine them to mint one $W$. This design parallels the creation/redemption mechanism of ETFs, where authorized participants swap baskets of underlying assets for ETF shares and vice versa\cite{Todorov2021}. Here, the role of authorized participant could be permissionless if the smart contract allows open access to the swap (subject to any regulatory whitelisting), or it could be restricted to certain parties if compliance requires it. In either case, the economic effect is that it ties the value of $W$ to the aggregate value of its constituents.

\subsection{Pricing and Arbitrage Mechanism}
Because of the enforced convertibility, the price of the everything token $P(W)$ in an efficient market will be tightly linked to the prices of the element tokens $P(E_1), \ldots, P(E_n)$. Specifically, under equilibrium conditions one would expect:
\begin{equation}
    P(W) \approx a_{1} P(E_{1}) + a_{2} P(E_{2}) + \cdots + a_{n} P(E_{n}),
    \label{eq:nav}
\end{equation}
which is analogous to the net asset value (NAV) of an ETF being the sum value of its holdings. If the market price of $W$ diverges from this computed benchmark, arbitrage opportunities arise. For instance, if $P(W)$ rises above the right-hand side of Eq.\,\eqref{eq:nav} (i.e., $W$ is overpriced relative to its components), an arbitrageur can purchase the requisite amounts of each element token $E_i$ on the open market and then redeem (or construct) one $W$ token from them via the smart contract, and finally sell the $W$ at the higher market price. This arbitrage process increases the supply of $W$ (putting downward pressure on $P(W)$) and increases demand for $E_i$ (pushing $P(E_i)$ up) until prices realign. Conversely, if $P(W)$ falls below the sum of its parts, an arbitrageur can buy one $W$ cheaply, redeem it to obtain $a_i$ of each $E_i$, and then sell those element tokens individually for a profit. This reduces the supply of $W$ (raising its price) and increases supply of $E_i$ (lowering their prices) until the gap closes. Through these mechanisms, the two-way convertibility ensures that $P(W)$ remains anchored to the intrinsic asset value as represented by the basket of element tokens, much like ETF arbitrage keeps fund prices aligned with underlying assets\cite{Todorov2021}.

The arbitrage mechanism not only stabilizes prices but also provides continuous feedback between micro-level component markets and the macro-level asset valuation. Each element token's price $P(E_i)$ is determined by its own supply-demand dynamics (which might include factors beyond the specific project, especially if the token is fungible across multiple projects or has an external commodity price link). For example, a surge in global copper prices would reflect in $P(E_{\text{Cu}})$ (a copper output token) rising; this would, via Eq.\,\eqref{eq:nav}, increase the implied fundamental value of a mining project's everything token that bundles copper output. Traders in the everything token market would then adjust $P(W)$ upward, or else face arbitrage. In this way, the architecture improves \emph{price discovery}: each component of value is transparently priced, and the whole asset's price becomes a composite that more accurately and dynamically reflects all underlying factors.

It is important to note that fees or slippage in conversions, as well as liquidity differences between $W$ and the $E_i$ markets, could create short-term deviations or transaction costs for arbitrageurs. However, these can be managed by protocol design (e.g., small minting/redemption fees to prevent abuse, or time-limited arbitrage windows) and are analogous to creation/redemption fees and bid-ask spreads in ETF markets. As long as the frictions are not too high, the core principle of arbitrage-driven alignment holds.

\subsection{Implementation Considerations}
In practice, implementing the element/everything token model requires careful attention to smart contract security, regulatory compliance, and governance. Smart contracts must reliably enforce the bundling ratios $a_i$ and manage the minting and burning of tokens. Standards such as ERC-1155 (multi-token standard) could be advantageous here, as they allow a single contract to manage multiple token types and their interactions\cite{Joshi2022}. Under an ERC-1155 implementation, one could assign different token IDs to $E_1, \ldots, E_n$ and another ID to $W$, and enforce through contract logic that transferring a $W$ out of the contract requires depositing the correct amounts of each $E_i$.

Regulatory classification of the tokens is another key consideration. The everything token $W$ effectively tokenizes an entire asset and would likely be deemed a security in many jurisdictions, meaning its issuance and trading would need to comply with securities laws (similar to security tokens in STOs). The element tokens, depending on their nature, might be viewed as commodity tokens, derivative contracts, or also securities if they represent claims on future revenue. The architecture allows flexibility: projects could be structured such that the $W$ token is offered in a regulated token offering for fundraising, while the $E_i$ tokens could be used internally or in B2B markets (for instance, energy tokens traded among utilities, or carbon tokens on carbon credit exchanges). A governance framework is needed to handle how and when new element tokens can be issued (e.g., if an asset increases production capacity) or how the token economics adjust if parts of the asset are sold or expire. These issues, while beyond the scope of our current discussion, must be addressed in real-world deployments. Encouragingly, initial empirical studies of tokenized assets emphasize that proper governance and regulatory adaptation are crucial to maintain market stability as these new models emerge\cite{Tanveer2025}.

\section{Use Case Illustrations}
To concretely demonstrate the element/everything token approach, we consider several types of assets and outline how they could be represented under this model. These examples span alternative assets, infrastructure, enterprise equity, environmental assets, and cross-border projects, highlighting the flexibility of the architecture.

\subsection{Alternative Assets (e.g., Mines, Fisheries)}Consider a gold-and-copper mining operation. Traditionally, an investor must buy a share of the entire mining company or project. Under our model, the mine could issue element tokens for its outputs and rights: e.g., an $E_{\text{Au}}$ token redeemable for a unit of gold produced, an $E_{\text{Cu}}$ token for copper output, a land-right token $E_{\text{land}}$ representing a hectare of mining concession, and an $E_{\text{permit}}$ token for the mining license rights. The everything token $W_{\text{Mine}}$ (e.g., "Copper-Gold Mine Token") would bundle a fixed proportion of these (reflecting, say, the expected life-of-mine output and rights per share of the project). An investor could either hold $W_{\text{Mine}}$ to get exposure to the entire project's performance or break it apart to separately trade the gold and copper tokens, which might be desirable if gold and copper prices diverge. Similarly, in a fisheries asset (such as a commercial fishing quota), element tokens could include quota rights for catch (tons of fish), tokens for cold storage capacity, and tokens for vessel usage or fuel. An everything token $W_{\text{Fishery}}$ would package these, allowing investors to trade the fishery as a whole or focus on specific components (e.g., selling the fish output forward via the quota token).
    
 \subsection{Infrastructure and Energy Projects} Infrastructure assets often have multiple value streams and associated credits. For example, a solar photovoltaic (PV) power plant generates electricity (measured in MWh), potentially earns renewable energy certificates or carbon credits, sits on land (land lease rights), and comprises equipment with residual value. We can define $E_{\text{energy}}$ tokens for each MWh of power production, $E_{\text{carbon}}$ tokens for certified carbon offsets from the plant's generation, $E_{\text{land-lease}}$ tokens for the land use per year, and $E_{\text{equip}}$ tokens representing claims on the salvage/residual value of the solar panels (or depreciation shares). The everything token $W_{\text{Solar}}$ (a "Solar Plant Token") then corresponds to a fixed bundle, say per year of operation it bundles 1000 MWh tokens, 1000 carbon tokens, the necessary land token, and equipment tokens. This enables separate trading of energy and carbon markets if desired; an investor primarily interested in energy can just hold $E_{\text{energy}}$, while another wanting the full project returns can hold $W_{\text{Solar}}$. A similar structure can be applied to a hydrogen production facility: element tokens for hydrogen output (kg of $\text{H}_2$), for storage capacity (tank volume units), and for any patented technology or emission credits, combined into a $W_{\text{H2Facility}}$ token representing the entire facility. Notably, such fractionalization of solar and other renewable projects has been cited as a way to attract wider investor participation and improve financing\cite{IdeaUsher2023}.
    
\subsection{Private Equity in Enterprises (Unlisted Companies)} Consider an agricultural processing plant (e.g., a factory that processes crops into food products). Its value comes from its physical assets (machinery, buildings), inventory (raw inputs and processed outputs), and operational rights or contracts. Element tokens could be defined for machine capacity or output (e.g., processing equipment hours or throughput tokens), for inventory units (tokens representing stored raw material or batches of finished goods that can be sold), and for property usage (a token for land or factory floor space rights). An $W_{\text{AgriPlant}}$ token would combine these such that one whole token equates to a proportional share of the factory's full capacity and assets. Another example is a data center: element tokens might include $E_{\text{rack}}$ (each token representing one server rack slot or a certain computing capacity), $E_{\text{bandwidth}}$ (network bandwidth allocation), $E_{\text{power}}$ (electricity consumption quota), and $E_{\text{carbon-offset}}$ (to neutralize emissions). An integrated $W_{\text{DataCenter}}$ token would package, say, 1 rack + X bandwidth + Y kWh + Z carbon offsets to represent a share of an operating data center. Investors could thus invest directly in specific facets of the business (for instance, buying more of the $E_{\text{rack}}$ if they believe computing demand will spike) or in the entire enterprise via the everything token.

\begin{table*}[t]
  \caption{Use-Case Map from Element Tokens to Everything Tokens}
  \label{tab:usecases}
  \centering
  \small
  \setlength{\tabcolsep}{6pt}
  \renewcommand{\arraystretch}{1.12}
  \begin{tabularx}{\textwidth}{@{}p{0.13\textwidth} p{0.15\textwidth} p{0.29\textwidth} p{0.36\textwidth} @{}}
    \toprule
    \textbf{Sector} & \textbf{Asset Example} &
    \textbf{Element Tokens (standardized, fully collateralized)} &
    \textbf{Everything Token (fixed bundle) \& Key Benefits} \\
    \midrule
    Alternative Resources & Copper--Gold Mine &
    Au output; Cu output; land / concession right; mining permit &
    W\_Mine $\Leftarrow$ Au + Cu + Land + Permit; \emph{Benefits:} sum-of-parts valuation; metals hedging \\
    \addlinespace[2pt]
    Fisheries & Industrial Fishery &
    catch quota; cold-chain storage right; vessel operation right &
    W\_Fishery $\Leftarrow$ Quota + Cold-chain + Ops; \emph{Benefits:} quota monetization; capex/ops financing \\
    \addlinespace[2pt]
    Infrastructure / Energy & Solar PV Plant &
    MWh energy; CO$_2$ credits; land lease-year; equipment depreciation / salvage &
    W\_Solar $\Leftarrow$ MWh + CO$_2$ + Land + Equip; \emph{Benefits:} unbundle revenue vs.\ ESG; green financing \\
    \addlinespace[2pt]
    Infrastructure / Energy & Hydrogen Facility &
    H$_2$ output (kg); storage capacity; IP / patent license &
    W\_H2 $\Leftarrow$ H$_2$ + Storage + IP; \emph{Benefits:} offtake prepay; tech/IP valuation \\
    \addlinespace[2pt]
    Private Equity & Agri-Processing Plant &
    equipment capacity; raw inventory; finished-goods inventory; land / factory use &
    W\_Agri $\Leftarrow$ Equip + Raw + Finished + Land; \emph{Benefits:} working-capital flexibility; inventory finance \\
    \addlinespace[2pt]
    Digital Infrastructure & Data Center &
    rack capacity; bandwidth; power quota (kWh); CO$_2$ offsets &
    W\_DC $\Leftarrow$ Rack + BW + Power + CO$_2$; \emph{Benefits:} capacity hedging; energy/carbon separation \\
    \addlinespace[2pt]
    Carbon / Environment & Carbon Offset Project &
    CO$_2$ credits; renewable energy certificates (REC) &
    W\_CO $\Leftarrow$ CO$_2$ + REC; \emph{Benefits:} compliance access; transparent retirements \\
    \addlinespace[2pt]
    Forestry & Forestry Carbon Sink &
    land tenure; timber harvest right; carbon sequestration credits &
    W\_Forest $\Leftarrow$ Land + Timber + CO$_2$; \emph{Benefits:} dual timber/carbon monetization; lifecycle pricing \\
    \addlinespace[2pt]
    Cross-Border Infrastructure & African Hydropower Project &
    MWh energy; land / water usage right; CO$_2$ credits; concession / grant &
    W\_Hydro $\Leftarrow$ MWh + Land/Water + CO$_2$ + Concession; \emph{Benefits:} concession-linked risk isolation; blended finance \\
    \bottomrule
  \end{tabularx}
\end{table*}

\subsection{Environmental and Carbon Assets} Projects like reforestation, carbon capture, or renewable energy credit programs yield intangible environmental assets that are increasingly traded. Under a single-token approach, one might issue a token that entitles the holder to all benefits of, say, a forest conservation project (timber, carbon credits, biodiversity credits). Using our model, a forestry project could issue $E_{\text{land}}$ tokens for land ownership or use rights, $E_{\text{timber}}$ tokens for permissible timber harvest volumes, and $E_{\text{carbon}}$ tokens for carbon sequestration credits (carbon offsets generated by the forest growth). The everything token $W_{\text{Forest}}$ (Forestry Carbon Sink Token) would combine these in the ratio that one token corresponds to, for example, one hectare of forest over a certain period, including its carbon and timber yields. Similarly, a carbon offset project (like a renewable energy installation feeding into a carbon market) might have separate tokens for carbon credits and for renewable energy certificates, with an overall project token linking them. This segmentation can enhance transparency for buyers who may only want the environmental attributes (carbon credits) separate from any financial returns of the project.
    
\subsection{Cross-Border or Multi-Jurisdiction Projects} Projects in emerging markets or involving public-private partnerships often involve additional layers of risk and rights. For instance, an African hydropower dam project could be tokenized by splitting: $E_{\text{power}}$ tokens for each unit of electricity generated, $E_{\text{water}}$ tokens for water usage rights or irrigation benefits, $E_{\text{carbon}}$ tokens for any carbon credits earned by clean energy, and $E_{\text{concession}}$ tokens representing the government concession or public-private partnership contract value. The composite $W_{\text{Hydro}}$ token would constitute an investment in the entire hydropower project combining those elements. Investors wary of certain country risks or regulatory changes could adjust their holdings accordingly—if, say, new regulations affect water rights, the $E_{\text{water}}$ token price might drop independently without dragging down the value of power generation tokens, allowing a more nuanced response than if all value were fused into a single indistinct asset.

These examples illustrate how the two-tier token model can be adapted to a wide range of use cases. The element tokens in each scenario are designed to be as granular as practical and economically meaningful, which not only aids in attracting specialized investors (e.g., commodity buyers interested in the gold output of a mine, or tech companies interested in data center capacity) but also facilitates cross-project and cross-sector trading of common elements. For instance, an $E_{\text{carbon}}$ token from the solar plant and one from the forestry project could be fungible if they both conform to a verified carbon credit standard, creating a larger unified market for carbon tokens across projects. This cross-cutting liquidity is a key advantage of standardizing element tokens.

\section{Discussion and Implications}
\subsection{Investor Perspective}
For investors, the element/everything token architecture offers unprecedented flexibility and transparency. By lowering the unit size of investment through fractional element tokens, it broadens access to asset classes that were previously out of reach\cite{Waliczek2025}. A retail investor could choose to invest in just the energy output of a renewable project or just the metal output of a mine, aligning their investment with their market outlook or hedging needs. The ability to partially exit a position is another advantage: rather than selling an entire project stake (which could be illiquid or take time to divest), an investor holding an everything token could redeem it and sell whichever element tokens they wish to reduce exposure to (for example, liquidating the carbon credit tokens if carbon prices are high, but keeping the energy tokens for long-term revenue). This greatly enhances \emph{liquidity} and allows dynamic portfolio rebalancing.

Moreover, transparent pricing of each asset facet makes valuation more straightforward and reduces information asymmetry. All investors can observe the market prices of the element tokens, which serve as price signals for the asset's components. This can lead to more accurate pricing of risk: if a particular component (say the regulatory permit token) carries high risk, its price will be discounted accordingly and that will reflect in the overall asset token only to the extent of that component's weight. In essence, investors can see exactly what they are paying for each part of the asset, unlike traditional bundled investments where opaque internal valuations may hide such details.

The architecture also enables \emph{cross-asset diversification at the component level}. An investor could assemble a portfolio of just copper output tokens from various mines around the world, effectively creating a diversified metals investment without the need to buy equity in mining companies. Or, they might hold an everything token for a wind farm plus extra carbon tokens from other projects to overweight the environmental attributes. This granular approach to investment could give rise to new strategies and derivatives (for example, one could imagine futures or options on individual element tokens like energy tokens, allowing hedging of specific risks).

It is worth noting that the complexity of managing multiple tokens might be a barrier for some investors. Asset management platforms and wallets would need to simplify the user experience, perhaps by providing a consolidated view or synthetic instruments for those who want a single exposure. However, the trend in decentralized finance is towards composability and user-friendly aggregation, suggesting that such tools would emerge as the market matures.

\subsection{Asset Owner Perspective}
For asset owners and project developers, the element/everything token model provides innovative financing and asset management capabilities. By issuing an everything token, a project can raise capital in a manner similar to issuing equity or project shares, but with the added appeal that investors know the token is directly backed by concrete asset components. The element tokens can be used to monetize specific asset outputs in advance; for instance, a mining company could sell some gold tokens forward to secure funding from gold purchasers, effectively pre-selling a portion of its production without giving away ownership of the entire mine. Meanwhile, it could retain or separately sell other element tokens (like copper or royalties) to different parties. This \emph{unbundling of asset value} allows tailoring the financing mix to those most interested in each component, potentially lowering the overall cost of capital compared to a one-size-fits-all investment offering.

Another advantage is ongoing liquidity and price feedback. Even if the project owner retains a significant portion of the tokens, the trading of some tokens on the open market provides continuous valuation benchmarks. The owner can mark-to-market the components of their asset and possibly make operational adjustments. For example, if the market values energy tokens from a solar farm much higher than the carbon tokens (indicating strong demand for energy but weaker for carbon credits), the project might choose to expand capacity or alter operations to maximize energy output relative to carbon generation. In a way, the market is signaling which aspect of the asset is more valuable.

From an operational standpoint, distributing revenue to token holders can be automated via smart contracts. Each element token could be set up to automatically receive its share of any corresponding revenues (for instance, if a utility pays the solar farm for electricity, that payment could trigger the smart contract to distribute stablecoin to each $E_{\text{energy}}$ token holder proportionally). The everything token would entitle holders to all streams, which can be achieved either by directly paying $W$ holders or by requiring conversion to elements to claim each stream separately. Different implementations are possible, but the result is that token holders can get \emph{programmable cash flows}—a feature not easily achievable with traditional equity without intermediaries.

Asset owners should also be cognizant of the regulatory implications. Splitting an asset into multiple tokens might trigger various regulatory regimes (securities law, commodities regulation, etc.) for different token types. Compliance can be maintained by whitelisting investors, embedding transfer restrictions, or registering tokens as necessary. The architecture itself is neutral to these choices, but in practice a permissioned or hybrid model might be employed for sensitive assets. Nonetheless, even traditional institutions like banks and stock exchanges are exploring tokenization precisely to handle such fractional ownership with compliance, suggesting that regulatory barriers can be overcome with collaboration\cite{Waliczek2025}.

\subsection{Market-Level Implications}
If adopted broadly, the element/everything token architecture could lead to a more interconnected and efficient market for Alternative Assets. By establishing a common framework to digitize and trade everything from infrastructure to natural resources, it creates a \emph{unified market infrastructure} akin to how stock markets aggregate corporate equity trading. Previously non-standard, bespoke transactions (like selling a portion of a power plant or leasing out mining rights) could migrate to standardized token markets with greater transparency and lower transaction costs. This not only unlocks value from previously illiquid assets but also could improve resource allocation in the economy: capital can more easily flow to asset components that yield the highest returns or fulfill demand (as evidenced by their token prices).

The pricing transparency and continuous arbitrage should reduce instances of mispricing or “trapped value.” For example, conglomerate discounts (where a company owning diverse assets trades at less than sum-of-parts value due to opacity) might be mitigated if each part has a token price that investors can point to. In essence, it fights the opacity and complexity that often plague large projects or companies by making them legible as a set of simple tokens.

On the flip side, the model introduces new complexities to manage. Markets for the element tokens could potentially be less liquid than the everything tokens if they attract only niche participants, which could lead to volatility. There's also the risk of partial markets: if some element tokens do not trade actively (say nobody trades the land-right token of a project), then the price discovery for that component may be weak, and $W$ pricing might rely on models rather than market prices for that piece. One mitigation is that market makers or the asset issuer could provide liquidity in all key components to ensure continuous pricing.

Another implication is the possibility of \emph{disaggregated ownership and control}. In a traditional asset, one owner (or a set of shareholders) has holistic control. With element tokens, different parties might effectively control different aspects (imagine one entity accumulates a majority of the land tokens, another accumulates energy tokens). This raises questions about governance: decisions about the asset (like expanding capacity or taking it offline) affect all components and thus all token holders. Governance tokens or agreements might be needed to coordinate token holder interests. This is an area for future exploration—perhaps a third token type (governance token) or using the everything token as the governance right while element tokens are non-voting claims.

Lastly, from a financial stability perspective, the architecture echoes some traits of structured finance and derivatives, which warrants careful risk assessment. The two-way convertibility and arbitrage closely tie markets together, which generally enhances stability by removing arbitrage gaps, but can also transmit shocks rapidly across markets. If one element token crashes in price (for instance, the carbon credits market collapses), it will immediately affect the everything token and by extension all holders. In that sense, token holders are still exposed to the full set of risks, but at least they are transparent and possibly hedgeable. Studies have noted that while tokenization can improve efficiency, it also introduces new interconnections and requires robust risk management\cite{Tanveer2025}. Policymakers might need to monitor these markets to ensure that fragmentation of assets into tokens does not lead to unforeseen systemic issues.

\subsection{Implementation Considerations: Smart Contracts, PoB, and Liquidity}
\label{sec:implementation}

Deploying the Element/Everything architecture requires an end-to-end stack that ties on-chain claims to verifiable off-chain production while preserving market microstructure properties (price alignment, liquidity, and composability). We outline a pragmatic blueprint.

\subsubsection{Smart-Contract Stack.}

\textbf{(i) Element Token Contracts.} Each standardized element $E_i$ is implemented as a fungible token with strict mint/burn controls. Contracts SHOULD expose pausability/allowlists for regulatory actions and emit granular events for custody and audit. Minting is gated by oracle attestations (Sec.~\ref{sec:pob-oracles}) to ensure that outstanding supply remains fully collateralized by the referenced real asset component. \textbf{(ii) Everything (Composite) Contract.} A per-asset contract specifies the composition vector $\mathbf{a}=(a_1,\dots,a_n)$ and enforces \emph{creation/redemption} at exact ratios: depositing $(a_1,\dots,a_n)$ of $(E_1,\dots,E_n)$ mints one $W$; burning one $W$ returns the basket. This makes $P(W)$ track $\sum_i a_i P(E_i)$ up to frictions via ETF-style arbitrage.\cite{Madhavan2016,PanZeng2017,Brown2021} Fractional issuance is handled by scaling $\mathbf{a}$, and all conversions MUST be evented for indexers and surveillance. \textbf{(iii) AIO and Yield Distribution.} A regulated primary offering (AIO) can be implemented as a sale contract that escrows the initial $W$ supply (or a separate wrapper) and releases tokens against payment schedules and lockups. A per-asset \emph{yield pool} contract accumulates distributable value and pays pro rata to token holders (pull-based claims or periodic streams). Streaming/epochal payout choices should consider gas economics and user custody; unclaimed yield may be auto-compounded into the pool subject to policy.

\subsubsection{\textbf{Proof-of-Behavior (PoB) and Oracle Layer}}
\label{sec:pob-oracles}
To anchor on-chain issuance to real productivity (e.g., MWh generated, ore tons mined), validators submit proofs of measured outputs; blocks or rewards are conditioned on accepted proofs. In practice, PoB integrates:
\begin{itemize}
  \item \emph{Data oracles:} Authenticated feeds from audited telemetry/IoT or third-party attestations. Trusted-execution and cryptographic designs (e.g., Town Crier, DECO) provide authenticated statements to contracts without trusted intermediaries.\cite{TownCrier2016,DECO2019}
  \item \emph{Verification and deviation control:} Multi-source aggregation, time-weighted averages, and deviation guards reduce manipulation; formal checks of oracle deviation risks (e.g., OVer) inform safe update policies and pause rules.\cite{OVer2024}
  \item \emph{Tokenization alignment:} Element minting hooks fire only upon validated output events; failed/contested proofs prevent inflationary drift of $E_i$ supplies.
\end{itemize}
PoB can run at L1 (protocol-native) or L2/off-chain with finalized attestations bridged on-chain; the economic requirement is identical: only verified behavior can expand tokenized claims.

\subsubsection{\textbf{Liquidity and Market-Making}}
Each $E_i$ pairs against a numeraire (e.g., stablecoin) in CFMM pools; basket NAV is synthesized by on-chain or off-chain routers. CFMM theory guarantees coherent pricing bands and convex trade optimality under mild assumptions,\cite{AngerisAFT2020,AngerisCFMM2021,AngerisUniswap2019} while the creation/redemption contract supplies the hard NAV boundary for $W$. Practical guidelines:
\begin{itemize}
  \item \emph{Dual surface liquidity:} Maintain depth on element pools and narrow conversion fees on $W$ to keep the arbitrage loop tight.
  \item \emph{Incentives:} LP rewards time-weighted by utilization/volatility prevent mercenary liquidity and support thin elements.
  \item \emph{Routing UX:} A meta-router chooses between (i) direct $W$ trade, (ii) redeem-then-sell-elements, or (iii) buy-elements-then-mint-$W$, returning best execution.
\end{itemize}

\section{Conclusion}
We presented a two-tier tokenization architecture that decomposes complex real-world assets into element tokens and an everything token, enabling both granular and aggregate trading of asset value. By bridging the gap between fractional and holistic ownership, this model addresses key liquidity and valuation challenges in assets such as infrastructure projects, natural resource ventures, and other traditionally illiquid investments. Element tokens create transparent markets for each component of an asset, while the everything token allows integrated exposure and serves as a vehicle for raising capital and distributing returns. The built-in convertibility and arbitrage mechanism link these markets to maintain coherent pricing, analogous to how creation/redemption keeps ETF prices aligned with underlying assets\cite{Todorov2021}. 

Our illustrative scenarios in mining, renewable energy, and other domains show that the architecture is broadly applicable and can be tailored to various industries. If implemented, it could unlock capital by making previously untradeable aspects of assets accessible to investors and by allowing asset owners to finance projects in novel ways. Investors benefit from lower barriers to participation, the ability to fine-tune their exposure, and improved price discovery, while asset owners gain flexibility in monetization and potentially more accurate market feedback on their assets. At the same time, the approach necessitates careful design of smart contracts and governance, as well as compliance with financial regulations. 

This work scratches the surface of the possibilities opened by asset tokenization when extended beyond a one-asset-one-token paradigm. Future research is warranted in several areas: (1) developing standardized taxonomies for element tokens across industries (to promote fungibility and comparability), (2) exploring automated market making and liquidity provision for interlinked token markets, (3) formalizing governance models for decentralized ownership of tokenized asset components, and (4) conducting case studies or pilots to empirically validate the benefits and identify any practical pitfalls. Additionally, collaboration with regulators will be important to ensure that such tokenized structures can operate within legal frameworks for securities, commodities, and property rights.

In conclusion, the element and everything token architecture provides a conceptual blueprint for how complex assets can be made more \emph{insightful} in terms of their value composition and more accessible as investable entities. By drawing inspiration from both traditional finance (securitization, ETFs) and decentralized finance (smart contracts, token standards), it represents a convergence of ideas aimed at one overarching goal: to make every asset, no matter how intricate, as easy to trade and invest in as a common stock or bond. Realizing this vision could mark a significant advance in the financialization of real assets and the efficiency of capital markets.

\bibliographystyle{IEEEtran}
\bibliography{references}

\end{document}